\documentstyle[prl,aps,epsfig]{revtex}
\tighten

\newcommand{\tr}{\mbox{trace}}

\newcommand{\ETAL}{{\it et al.}}

\newcommand{\TOT}{{\rm tot}}

\newcommand{\MAT}{{\rm m}}

\newcommand{\CDM}{{\rm c}}
\newcommand{\BAR}{{\rm b}}
\newcommand{\NEUT}{{\rm \nu}}
\newcommand{\PHOT}{{\rm \gamma}}

\newcommand{\SCAL}{{\rm S}}

\newcommand{\UUNIT}[2]{
{\;\mathrm{#1}^{#2}}
}

\newcommand{\MIX}{{\rm MIX}} 
\newcommand{\AD}{{\rm AD}}
\newcommand{\BI}{{\rm BI}} 
\newcommand{\CI}{{\rm CI}}
\newcommand{\NID}{{\rm NID}} 
\newcommand{\NIV}{{\rm NIV}}

\begin{document}
\draft
\twocolumn[\hsize\textwidth\columnwidth\hsize\csname@twocolumnfalse\endcsname

\title{Cosmic Microwave Background
       anisotropies with mixed isocurvature perturbations}

\author{R.~Trotta, A.~Riazuelo and R.~Durrer} 

\address{D\'epartement de Physique Th\'eorique, Universit\'e de
  Gen\`eve, 24 quai Ernest Ansermet, CH-1211 Gen\`eve 4, Switzerland}

\date{19 July 2001}

\maketitle

\begin{abstract}
  Recently high quality data of the cosmic microwave background
  anisotropies have been published.  In this work we study to which
  extent the cosmological parameters determined by using this data
  depend on assumptions about the initial conditions.  We show that
  for generic initial conditions, not only the best fit values are
  very different but, and this is our main result, the allowed
  parameter range enlarges dramatically.
\end{abstract}
]

\section{Introduction}

The discovery of anisotropies in the cosmic microwave background (CMB)
by the COBE satellite in 1992~\cite{COBE} has stimulated an enormous
activity in this field which has culminated recently with the high
precision data of the BOOMERanG~\cite{Net}, DASI~\cite{DASI} and
MAXIMA-1~\cite{MAX2} experiments.  The CMB is developing into the most
important observational tool to study the early Universe. So far, this
data have however mainly been used to determine cosmological
parameters for a fixed model of initial fluctuations, namely scale
invariant adiabatic
perturbations~\cite{Bal,jaffe,tegman,tegman2,lange,Pryke,Sto,deBe}.

In this work we investigate to which extent the determination of 
cosmological parameters depends on initial conditions. Rather than
exploring the most general parameter space, we mainly want to show in
a specific example how the allowed parameter range is enlarged when
the requirement for purely adiabatic initial conditions is loosened.
Therefore, in order to limit the computational effort we have chosen
to vary some cosmological parameters and keep the others fixed.  We
have set the total density parameter $\Omega_\TOT = \Omega_\Lambda +
\Omega_\MAT = 1$ and fixed $\Omega_\Lambda = 0.7$ and $\Omega_\MAT =
\Omega_\CDM + \Omega_\BAR = 0.3$. Here $\Omega_\Lambda$ denotes the
density parameter due to a cosmological constant, $\Omega_\Lambda =
\Lambda / 3 H_0^2$, while $\Omega_\CDM$ and $\Omega_\BAR$ are the
density parameters of cold dark matter (CDM) and baryons respectively,
and $H_0 = 100 h \UUNIT{km}{} \UUNIT{s}{-1} \UUNIT{Mpc}{-1}$ is the
Hubble parameter today. For fixed $\Omega_\Lambda$, $\Omega_\CDM$ and
spectral index $n_\SCAL = 1$ we determine the parameters $h$ and
$\omega_\BAR \equiv \Omega_\BAR h^2$ for generic initial conditions.

\section{Initial conditions}

A study considering the adiabatic together with just one isocurvature
mode has been undertaken recently~\cite{Luca}. To choose an even more
generic set of initial conditions we follow the procedure outlined in
Ref.~\cite{BMT1}.  The matter components present in the ``standard''
universe are CDM, baryons, massless neutrinos ($\nu$) and photons
($\PHOT$). Usually adiabatic initial conditions are assumed for the
perturbations of these components. In Newtonian or flat-slicing gauge,
this implies
\[ 
\Delta_\CDM
 = \Delta_\BAR
 = \frac{4}{3} \Delta_\PHOT
 = \frac{4}{3} \Delta_\NEUT , 
\mbox{  and  } 
V_\CDM
 = V_\BAR
 = V_\PHOT
 = V_\NEUT , 
\] 
where $\Delta_i$ and $V_i$ are the density fluctuation and the
peculiar velocity potential of component $i$.  However, this does not
represent the most general set of possible initial conditions for
cosmological perturbations.  Apart from the adiabatic mode given above
there are a baryon isocurvature mode ($\BI$) a CDM isocurvature mode
($\CI$) a neutrino isocurvature density mode ($\NID$) and a neutrino
isocurvature velocity mode ($\NIV$). The precise definition of these
modes is given in Ref.~\cite{BMT1} and some observable consequences of
deviations from a pure adiabatic model were already investigated in
Ref.~\cite{lr99}. The most generic initial conditions are then given
by a positive definite $5 \times 5$ matrix $M$ representing the
amplitude of each of these modes, including all the possible
cross-correlations (see Ref.~\cite{BMT1}; a gauge-invariant
formulation of the initial conditions is given in Ref.~\cite{Trotta}).
We have noticed that implementing the initial conditions for all the
modes in synchronous gauge was somewhat tricky, while this procedure
 is very simple and numerically unproblematic in
gauge-invariant formalism.

For a fixed set of cosmological parameters we first compute the CMB
anisotropy spectrum $C^{ij}_\ell$ when only one of the elements of the
correlation matrix is non-zero ($M_{ij} = 1$, all other elements
vanish) with a fixed spectral index $n_\SCAL = 1$ for all modes. Next
we set
\begin{equation}
\label{Cl}
C_\ell (M) = \sum_{i, j = 1}^5 M_{ij} C^{ij}_\ell ~.
\end{equation} 
As already noticed in Ref.~\cite{BMT2}, the $\BI$ and $\CI$ components
of the correlation matrix are identical, up to an irrelevant
multiplicative constant (which is due to the fact that $\Omega_\BAR
\neq \Omega_\CDM$). We have therefore restricted our analysis to the
four modes $\AD$, $\CI$, $\NID$, $\NIV$ without loss of generality.
We then vary the correlation matrix $M$ and the cosmological
parameters $\omega_\BAR$ and $h$ to search for the best fit to the
data using a maximum likelihood method.
\begin{figure}  
  \centerline{ \psfig{file=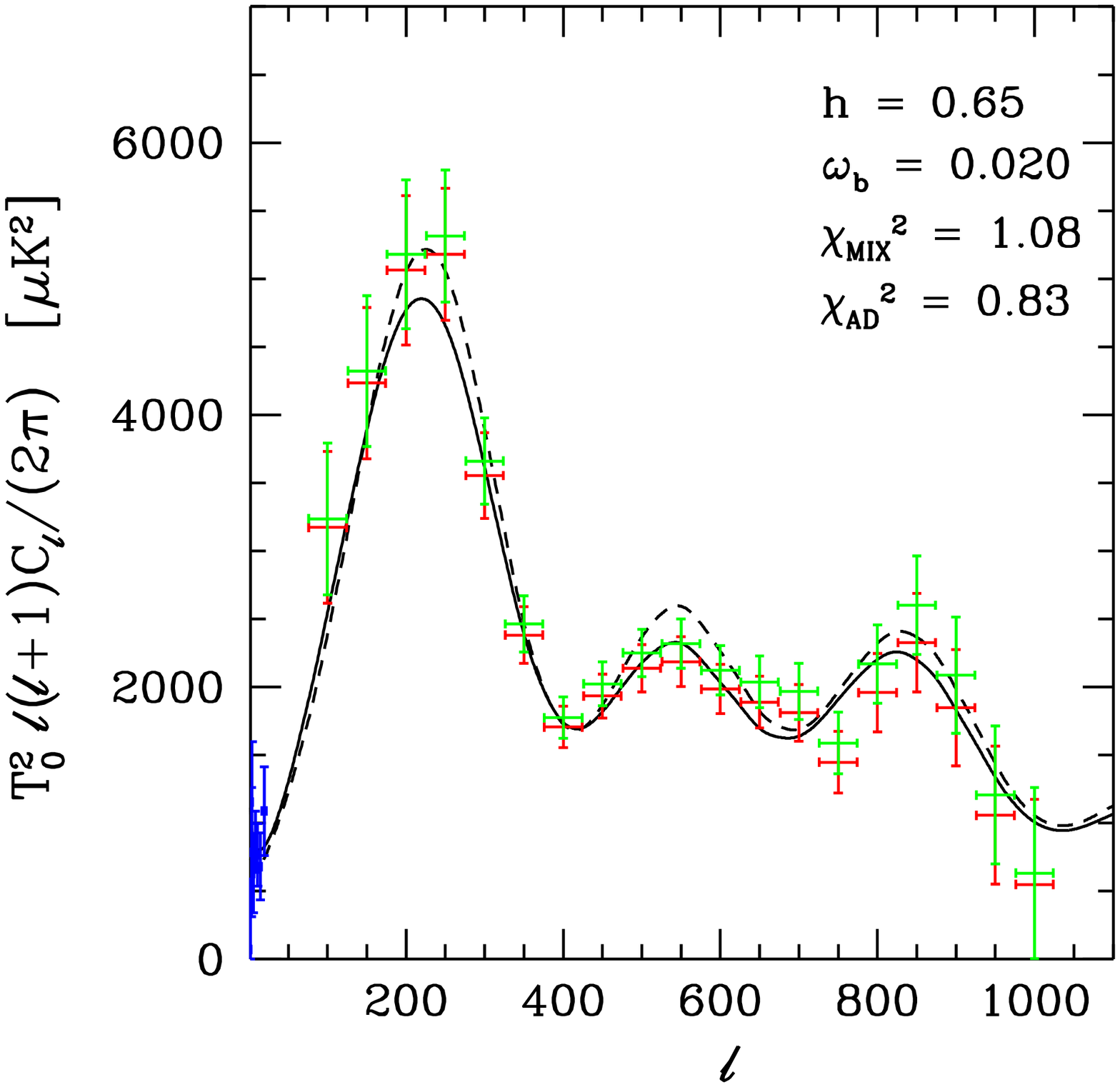,angle=0,width=3.5in}}
  \centerline{ \psfig{file=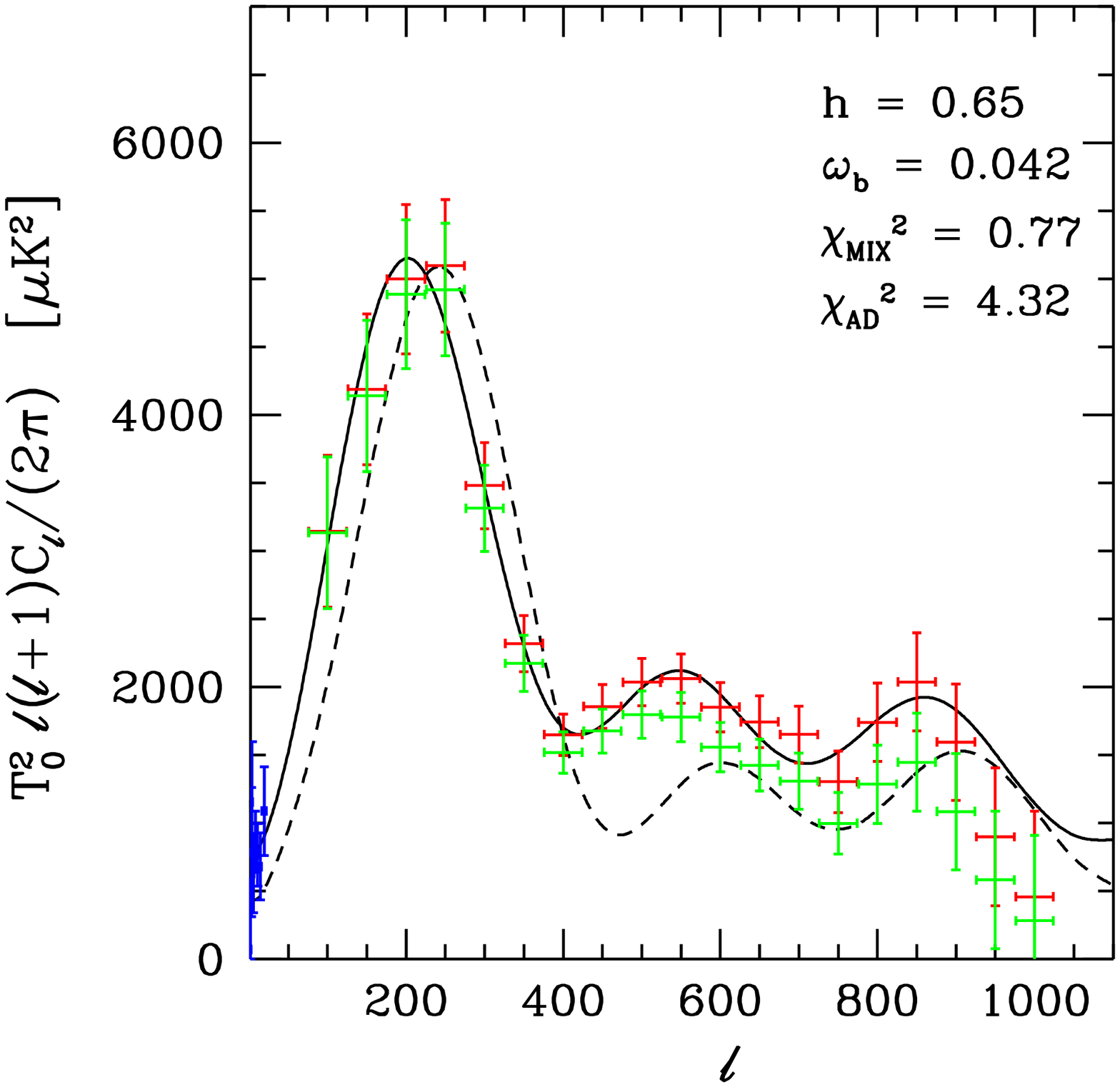,angle=0,width=3.5in}}
\caption{CMB anisotropy spectrum for different values of the 
  cosmological parameters $\omega_\BAR$ and $h$. We have shown the
  best-fit corresponding to a purely adiabatic case (dashed line) and
  allowing the most general set of initial conditions (solid line).
  The calibration and the beam size of the BOOMERanG data have been
  optimized to fit the mixed model (red/dark error bars) or the
  adiabatic model (green/light error bars). The parameter choice on top
  ($\omega_\BAR = 0.02$, $h = 0.65$) can be fitted by both models
  while the values $\omega_\BAR = 0.042$, $h = 0.65$, can only be
  fitted by a mixed model.  }
\label{spec}
\end{figure}  

Since the main point of this paper is to clarify the role of initial
conditions and not to obtain the most realistic set of cosmological
parameters, we restrict our analysis to the COBE\cite{COBE} and
BOOMERanG~\cite{Net} data. For the BOOMERanG data we also take into
account the calibration and the beam size uncertainties~\cite{Net}
which we treat just like two additional (normally distributed)
parameters of the problem.  The best fits are computed using a
downhill simplex method~\cite{nr} initiated after choosing a starting
point randomly.  Although this is certainly not the best method it
proved to be sufficient for our purpose. In practice, each fit is the
best fit among 15,000 minimization runs with random initial
conditions. The positive semi-definiteness of the correlation matrix
$M$ is insured by penalty functions on the sub-determinants of $M$
(more details are given in~\cite{Trotta}). It turns out that the
topology of the $\chi^2$ surface on our $14$-dimensional parameter
space is quite complicated with many local minima and probably many
degeneracies (see also the example discussed in~\cite{Luca}).

In Fig.~\ref{spec} we show the best fit spectra for two different
choices of the cosmological parameters $\omega_\BAR$ and $h$.  Both of
them are good fits if we allow for mixed initial conditions.  On the
plot we have also indicated the reduced $\chi^2$. For a fixed choice
of the parameters $\omega_\BAR$, $h$ the purely adiabatic model has
only $3$ parameters (the amplitude of the adiabatic mode, the
BOOMERanG calibration and beam size).  With 26 data points (7 from
COBE and 19 from BOOMERanG) this leads to $F_\AD = 26 - 3 = 23$
degrees of freedom, while the mixed models have a symmetric $4 \times
4$ matrix determining the initial amplitude, leading to a total of
$12$ parameters and hence only $F_\MIX = 14$ degrees of freedom. If we
also vary $\omega_\BAR$ and $h$, the number of degrees of freedom
is lowered by $2$. It is of course not surprising that for the
parameters $h = 0.65$, $\omega_\BAR = 0.02$, which are well fitted by
the adiabatic model, the reduced $\chi^2$ of the adiabatic model is
somewhat lower than the one of the mixed model. For the mixed model,
the absolute $\chi^2$ is always lower than the one of the adiabatic
model since the latter one is contained in the generic class of mixed
models, but the reduced $\chi^2$ can of course be higher since it is 
divided by a smaller number of degrees of freedom, $F_\MIX<F_\AD$.
This is actually the case for the parameters choice in the top panel 
of Fig.~\ref{spec}.

\begin{figure}[ht]
  \centerline{
    \psfig{file=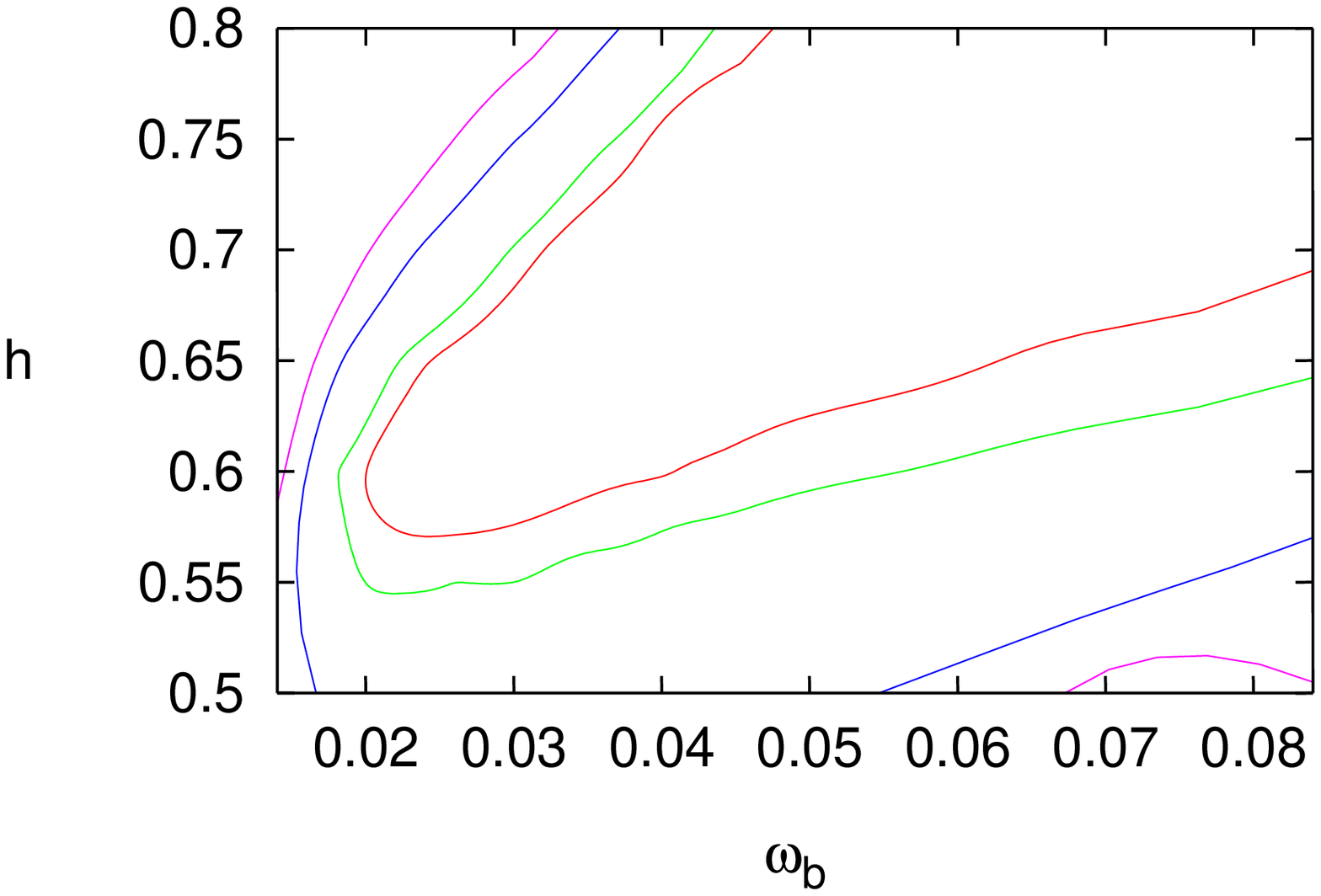,
      bbllx=150pt,bblly=130pt,bburx=490pt,bbury=610pt,
      angle=270,width=3.5in}}
    \centerline{
      \psfig{file=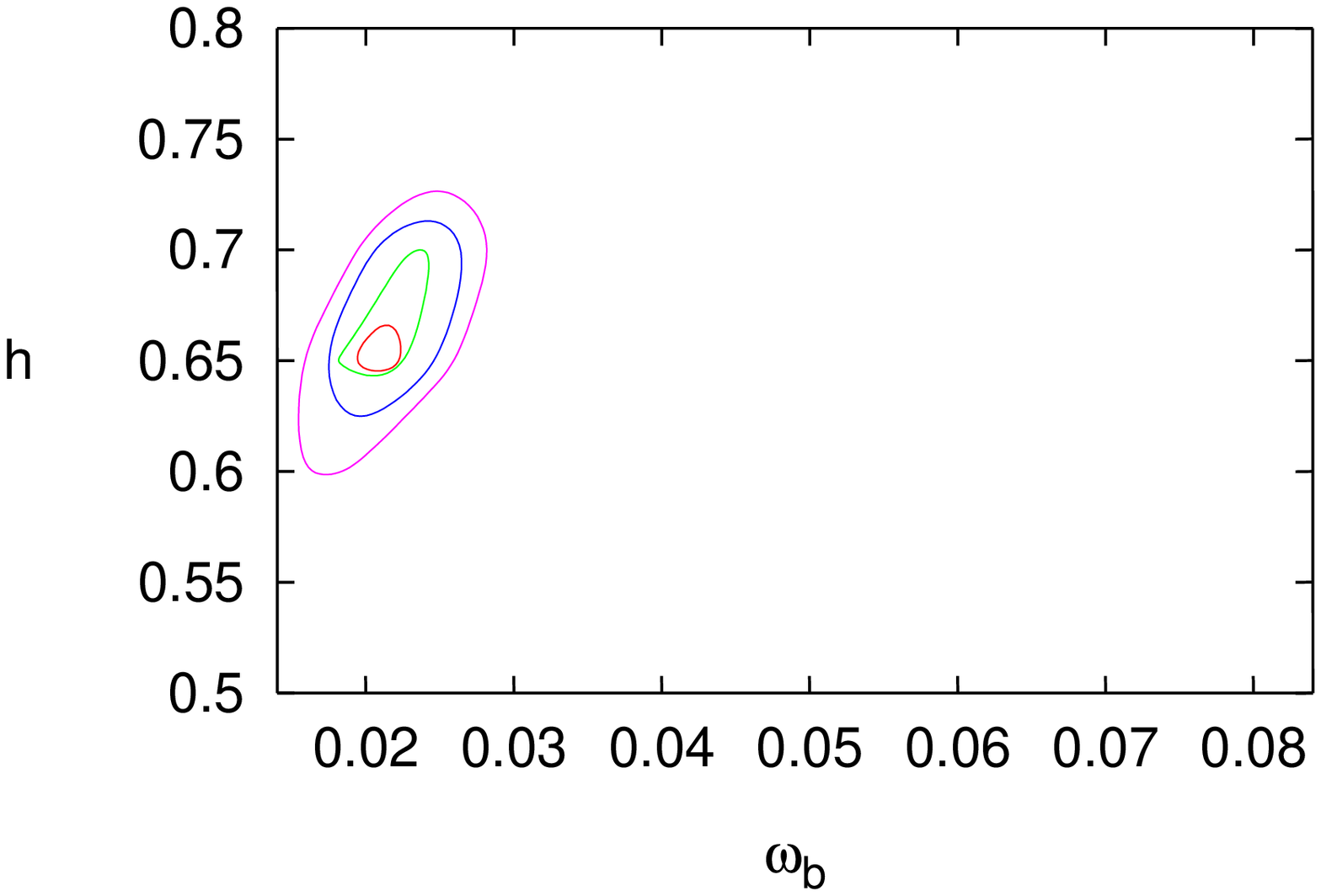,
        bbllx=150pt,bblly=130pt,bburx=490pt,bbury=610pt,
        angle=270,width=3.5in}}
\caption{The likelihood contours of 50\%, 68\%, 95\%, 99\%
  are indicated in the $(\omega_\BAR, h)$ plane for mixed models (top)
  and for purely adiabatic models (bottom). The likelihoods are
  obtained by marginalization over the BOOMERanG calibration and beam
  size as well as over the initial conditions given by the matrix $M$
  for mixed models and by the amplitude of the adiabatic mode for
  adiabatic models.  For the mixed model, the lowest $\chi^2$
  corresponds to even higher values of $\omega_\BAR$ and $h$ than
  those shown in the plot.}
\label{like}
\end{figure}

To indicate what happens when models with mixed initial conditions are
admitted we determine the likelihood functions of the cosmological
parameters $\omega_\BAR$ and $h$ by marginalizing over the initial
conditions and the BOOMERanG calibration and beam size. The result is
shown in Fig.~\ref{like} where the likelihood contours in the
$(\omega_\BAR, h)$ plane for likelihoods of 50\%, 68\%, 95\%, 99\% are
indicated for mixed models (top) and for purely adiabatic models
(bottom). This figure is our main result.
It is quite amazing to see to which extent the innermost good fit
contour opens up once we allow for isocurvature components.
Strangely, the only excluded region which remains is the upper left
corner corresponding to the value of $\omega_\BAR$ inferred from big
bang nucleosynthesis~\cite{burles} and the Hubble
space telescope key project value for the Hubble parameter~\cite{HST}
of $h = 0.72\pm 0.08$. 

There is absolutely no upper limit for $\omega_\BAR$ within the regime
investigated here! Actually, this can be understood by noting that
with adiabatic initial conditions, the strongest feature of a high
baryon density universe is the asymmetry between even and odd acoustic
peaks, but recalling that this asymmetry exists only in the matter
dominated era and not in the radiation dominated era. With the
parameters we have chosen here, the high value of $\Omega_\Lambda$
implies a low matter content and hence, equivalence between radiation
and matter happens not long before decoupling. For high values of $h$,
the equivalence occurs earlier and the asymmetry of the peaks allows
to put upper limits on the baryon density in purely adiabatic models.
The second feature of high baryon density is to reduce the sound
horizon and therefore to shift the acoustic peaks to the right. These
two effects constrain the baryon density most strongly in the class of
purely adiabatic models. In the case of mixed initial conditions, it
is known that the presence of the isocurvature modes, especially the
neutrino isocurvature modes, can not only shift the peaks
significantly but also change their relative heights~\cite{BMT2} (see
lower panel of Fig.~\ref{spec}). A high baryon density can therefore
easily be accommodated in this framework. Note however that for high
$\omega_\BAR$ and low $h$, it is difficult to find a good fit, even
for mixed models, because there is not enough power in the secondary
peak region. The reason is that the early integrated Sachs-Wolfe
effect boosts the first peak.

\begin{figure}[ht]
    \centerline{ 
      \psfig{file=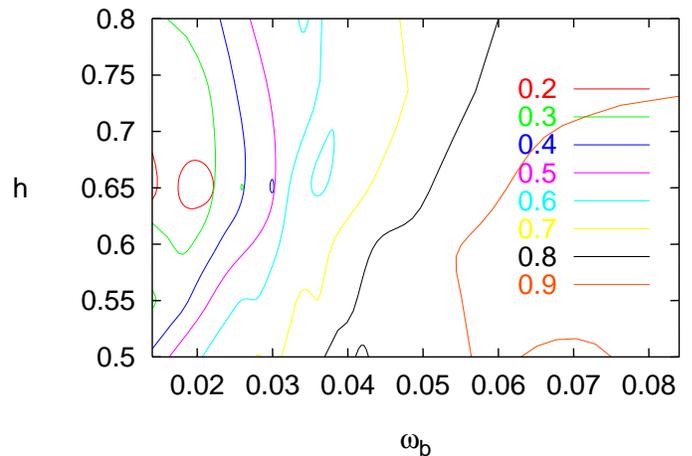,
        bbllx=150pt,bblly=130pt,bburx=490pt,bbury=610pt,
        angle=270,width=3.5in}}
  \caption{The isocurvature content $\alpha$ of the best fit mixed model at 
    fixed parameter values $(\omega_\BAR, h)$ is indicated. The
    contours $\alpha = 0.2$ to $0.9$ in steps of $0.1$ are shown.  }
\label{alpha}
\end{figure}

We define the isocurvature content of a mixed model by $\alpha =
(M_{22} + M_{33} + M_{44}) / \tr M$, where $M_{11}$ denotes the
adiabatic mode amplitude.  The isocurvature content in the good fit
model shown in the top panel of Fig.~\ref{spec} is only $\alpha = 0.12$,
while for the parameter choice in the bottom panel $\alpha = 0.69$.
Hence, if the cosmological parameters are close to those chosen in the
top panel, we can conclude that the cosmic perturbations are 
predominantly adiabatic.  In Fig.~\ref{alpha} we show the isocurvature
content $\alpha$ of the best fit model obtained by minimizing $\chi^2$
by variation of the initial conditions at fixed values of the cosmological 
parameters. Clearly, the further we move away from the
parameter region well fitted by the purely adiabatic model, the higher
becomes the isocurvature contribution needed to fit the data.

The main non-adiabatic component of our best fits is the $\NID$ mode.
This was to be expected as this mode and its correlator with the
adiabatic mode can shift the peak positions and can substantially add
or subtract from the second peak.  It is known that for interacting
species the non adiabatic part of the perturbations tends to decay
with time.  Therefore the generation of this $\NID$ component can only
occur after neutrino decoupling, that is at $T \lesssim 1
\UUNIT{MeV}{}$. Whether or not such a phenomenon can occur at low
energy is an open question. However, a neutrino isocurvature
perturbation can also be due to a fourth species of sterile neutrinos
which may have decoupled very early in the history of the Universe.
The same remark applies of course also to the CDM isocurvature mode.
Note that the energy density of this fourth neutrino type cannot be
very high in order not to contradict the light element abundances, but
there is nothing which prevents (at least in principle) the presence
of large perturbations in this fluid.

\section{Conclusion}

We have shown that allowing for isocurvature contributions one may
very well fit present CMB data with a set of cosmological parameters
which differs considerably from the one preferred by adiabatic
perturbations. More important, allowing for generic initial
conditions, the ranges of cosmological parameters which can fit the
CMB anisotropy data widen up to an extent to become nearly
meaningless.

On the other hand, assuming measurements of cosmological parameters
from other methods like direct measurements of the Hubble parameter
which yield $h\sim 0.65$ and big bang nucleosynthesis which implies
$\omega_\BAR \sim 0.02$, we can use the CMB to limit the isocurvature
contribution in the initial conditions (or other unconventional
features) and thereby learn something about the very early universe,
the inflationary phase which probably has generated these initial
conditions.

We believe that it is already interesting to note that for
cosmological parameters in the range preferred by other, CMB
independent, measurements ($\Omega_\Lambda \sim 0.7$, $\Omega_\MAT \sim
0.3$, $h \sim 0.65$, $\omega_\BAR \sim 0.02$) the isocurvature
contribution in the initial conditions has to be relatively modest.

Finally, our work shows the danger of calling parameter estimation by
CMB anisotropy experiments a ``parameter measurement'' since the
results depend so sensitively on the underlying model assumptions. We
rather consider CMB anisotropies as an excellent tool to test model
assumptions or consistency.  In the light of these findings, the
importance of non-CMB measurements of cosmological parameters can
clearly not be overstated. In short, CMB seems to be a better tool to
investigate the {\it primordial} parameters (i.e., the initial
conditions) rather than the {\it cosmological} parameters ($\Omega$'s,
$h$ etc).  Hopefully, measuring the CMB polarization will represent an
additional non-trivial mean to remove the degeneracy present between
cosmological parameters and initial conditions. This has been studied
in detail in Ref.~\cite{BMT2}. The main reason for this is that
polarization, which is induced only during decoupling, is mostly
sensitive to the photon dipole $V_\PHOT$ rather than the photon
density perturbation $\Delta_\PHOT$, these two quantities depending in
a different way on the initial conditions. In the same vein, using the
normalization of the matter power spectrum (provided it can be
measured accurately...) also helps to break some of the degeneracies
induced by the isocurvature modes.

\acknowledgments

We thank Kavilan Moodley, Martin Bucher, Neil Turok, David Langlois
and Bohdan Novosyadlyj for stimulating discussions.  This work is
supported by the Swiss National Science Foundation and by the European
network CMBNET.

\end{document}